\documentclass{iaus}

\usepackage{graphicx}

\title[Stellar variability and planetary transits] %% give here short title to appear on odd pages%%
{Reconstruction of the transit signal in the presence of stellar variability}%% give here main title to appear on front page%%

\author[Aude Alapini \& Suzanne Aigrain]   %% give here short author list to appear on even pages%%
{Aude Alapini$^1$ and Suzanne Aigrain$^1$} %% give here main author list to appear on front page%%

\affiliation{$^1$Astrophysics Group, School of Physics, University of Exeter, \\ Stocker Road, Exeter EX4 4QL, United Kingdom \\email: {\tt alapini@astro.ex.ac.uk},\ {\tt suz@astro.ex.ac.uk}}

\pubyear{2008}
\volume{249}  %% insert here IAU Symposium No.
\pagerange{1--4}
\jname{Exoplanets: Detection, Formation and Dynamics}
\editors{A.C. Editor, B.D. Editor \& C.E. Editor, eds.}

\begin{document}

\maketitle

\begin{abstract}
Intrinsic stellar variability can hinder the detection of shallow transits, particularly in space-based data. Therefore, this variability has to be filtered out before running the transit search. Unfortunately, filtering out the low frequency signal of the stellar variability also modifies the transit shape. This results in errors in the measured transit depth and duration used to derive the planet radius, and orbital inclination. We present an evaluation of the magnitude of this effect based on 20 simulated light curves from the CoRoT blind exercise 2 (BT2). We then present an iterative filter which uses the strictly periodic nature of the transits to separate them from other forms of variability, so as to recover the original transit shape before deriving the planet parameters. On average with this filter, we improve the estimation of the transit depth and duration by 15\% and 10\% respectively.  
 
\keywords{exoplanet, transit, stellar variability}
%% add here a maximum of 10 keywords, to be taken form the file <Keywords.txt>
\end{abstract}

\firstsection % if your document starts with a section,
              % remove some space above using this command.

\section{Introduction}

	\subsection{Planet parameters from observations and associated errors}

The radius ($R_{\rm p}$, eq. \ref{eq1}) and the mass ($M_{\rm p}$, eq. \ref{eq2}) of an exoplanet can be fully solved when measuring both the flux and the radial velocity variations of the parent star due to its orbiting planetary companion.
 
\begin{equation}
R_\textrm{p}=R_{\star} \sqrt{\frac{\Delta F}{F}}
\label{eq1}
\end{equation}

\begin{equation}
M_\textrm{p}=M_{\star}^{\frac{2}{3}}\frac{K}{\sin{i}}\left(\frac{P}{4\pi G}\right)^{\frac{1}{3}}
\label{eq2}
\end{equation}

\noindent where, $R_{\star}$ and $M_{\star}$ are the radius and the mass of the parent star, $\frac{\Delta F}{F}$ the flux variation due to the planet transiting the disc of its parent star, $K$ the amplitude of the radial velocity variation of the parent star due the gravitational influence of its orbiting planet, $i$ and $P$ the orbital inclination and period of the planet, and $G$ the gravitational constant. $\frac{\Delta F}{F}$, $P$, and  $i$ can be measured from the light curve. A common way to measure $R_{\star}$ and $M_{\star}$ is by comparing the stellar spectrum to stellar atmosphere models, allowing to derive the stellar parameters ($T_{eff}$,  $\log{g}$, [Fe/H]) used to obtain the stellar mass and radius. $R_{\star}$ can also be measured more precisely, and without the use of models, with interferometry, or with transit fitting (in the case of high precision light curves).

	\subsection{Planet parameters and planet evolution and formation models}

Improving the precision on observational planet masses and radii is important for both planet structure and planet formation models. The internal structure of a planet can be studied by comparing its mass and radius to model predictions of planets with different composition. Determining planet structure is important to derive observational statistics on planet types, which can then be compared to the predictions of planet formation models. \cite{SKHal07} show that to determinate the composition of sub-Uranus planets, error bars of 2\% on the planet parameters are required. The current uncertainties on planet masses and radii are of the order of 10\%. Improving these measurements is thus vital to help confirm the models.
 
	\subsection{Sources of uncertainties on planet parameters}

The uncertainties on the planet mass, radius, and inclination depend on the uncertainties on the host star mass and radius, on the uncertainties on the transit parameters ($\frac{\Delta F}{F}$, total transit duration), and on the uncertainties on the radial velocity measurements. For large planets ($>$Jupiter), the uncertainties on the planet mass and radius are mainly due to the uncertainties on the stellar parameters. For sub-Uranus planets around active stars, the uncertainties on the planet mass and radius can be dominated by the uncertainties on the transit parameters. 

\section{Side effects of pre-detection stellar variability filters \label{sec2}}

	\subsection{Pre-detection stellar variability filter}

Active stars shows intrinsic flux variations due to temporal evolution and rotational modulation of structures on their surface (stellar spots, plages, granulation). Intrinsic stellar variations can have amplitudes much greater than transits, and thus, can hinder transit detection. These variations occur at a lower frequency (longer time length) than transit events (\cite[chap. 3, fig. 3.2]{Sth05}). Pre-detection stellar variability filters use this difference to separate variations typical of stellar variability, from those on the time scale of transits (minutes to hours). \\

\begin{figure}[htbp]
\center
	 \begin{tabular}{cc}
	 \includegraphics[width=6cm]{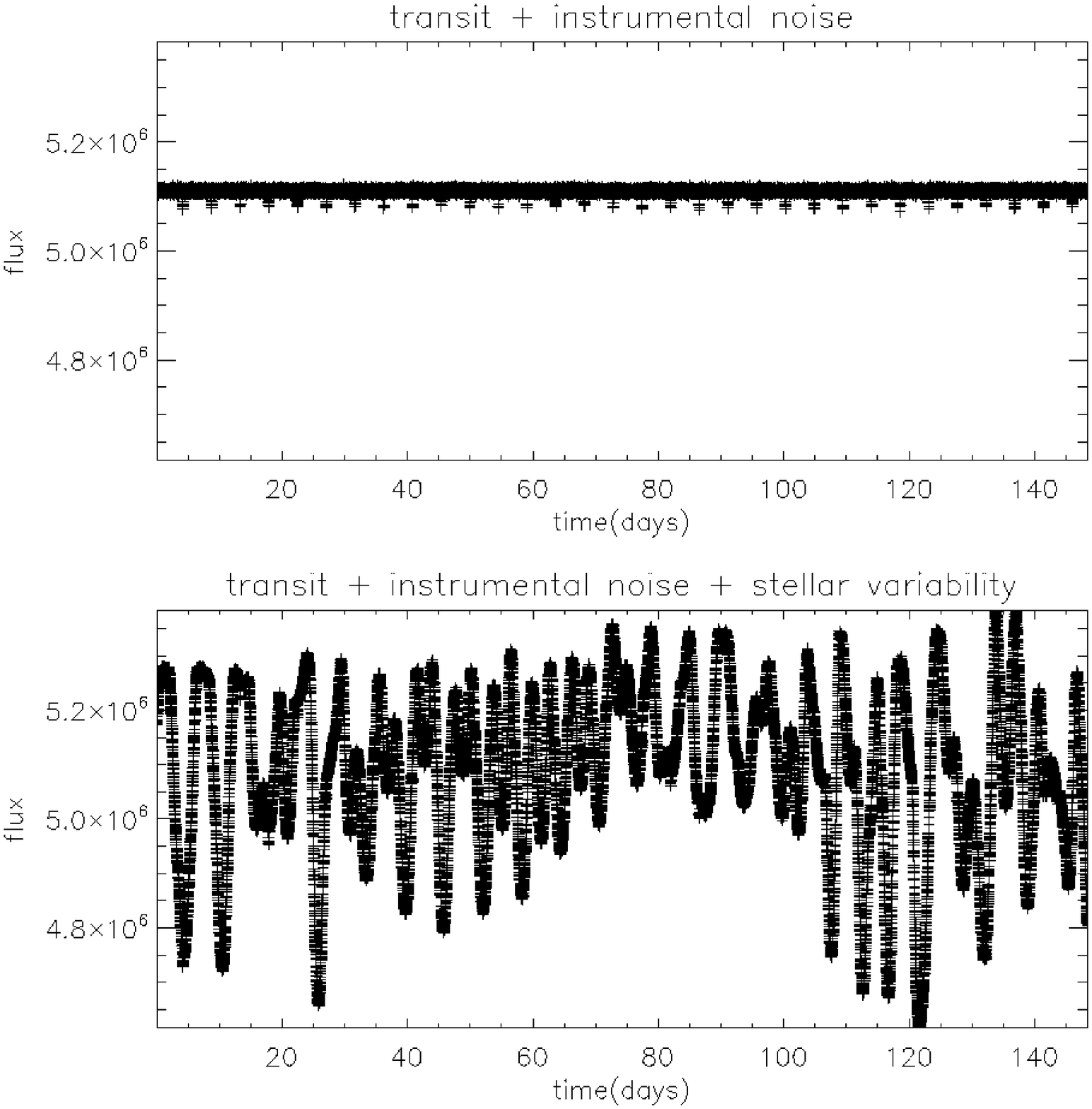}&\includegraphics[width=6cm]{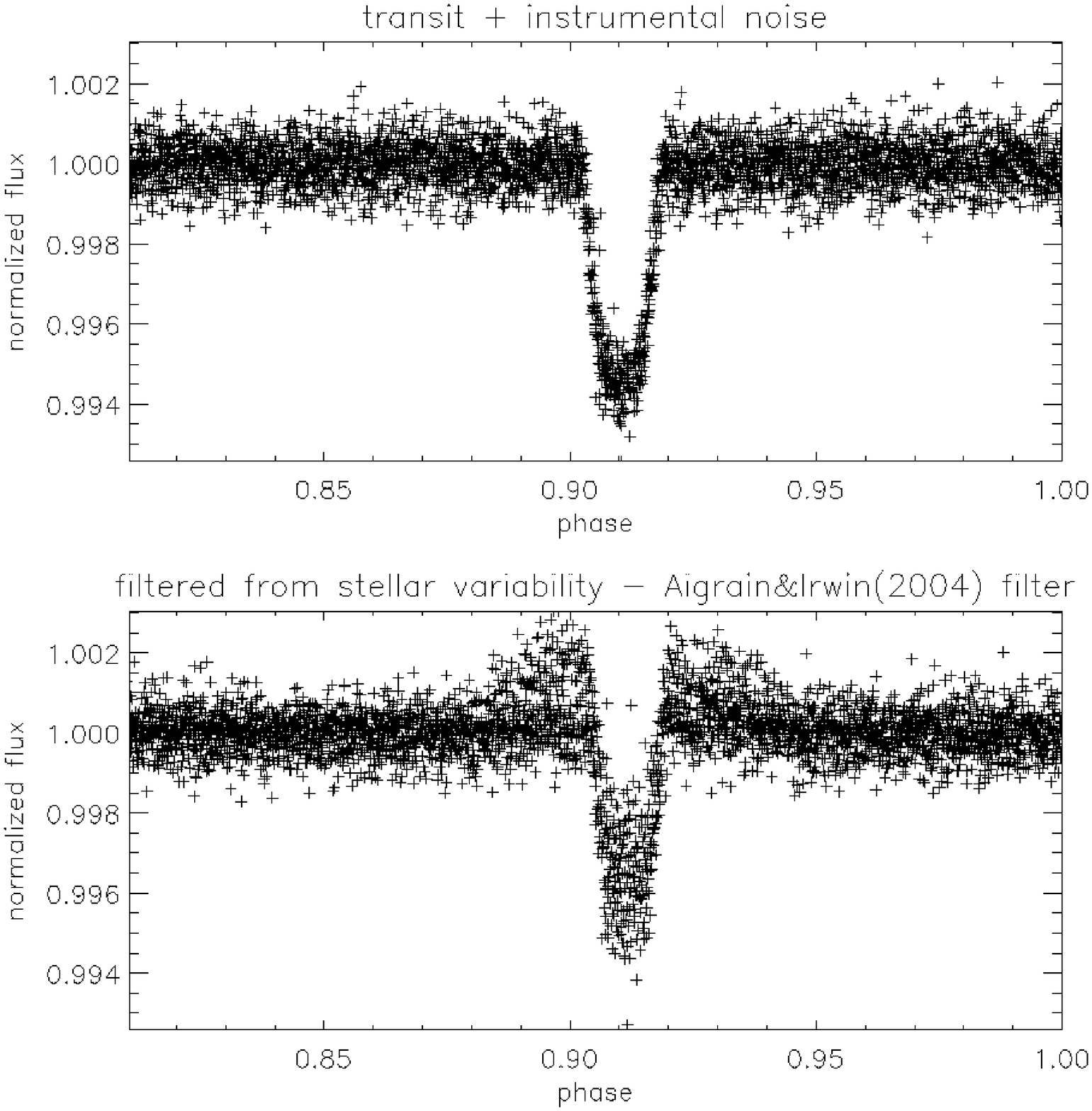}\\
	a & b \\
  	\end{tabular}
\caption{Stellar variability hinders transit detection (\underline{\it column a}): the {\it bottom plot} is an example of a CoRoT BT2 light curve with Jupiter like transits on an active host star. This light curve is composed of a simulated transit signal ({\it top plot}), the instrumental noise expected in CoRoT data, and simulated stellar variability. Pre-detection filters can deform the shape of transits (\underline{\it column b}): the {\it top plot} of {\it column b} is the phase fold of the {\it top plot} of {\it column a}, centered on the transit. The {\it bottom plot} of {\it column b} is the phase fold of the {\it bottom plot} of {\it column a}, filtered from stellar variability using \cite{AI04} filter.}
\label{fig1}
\end{figure}

The work presented in this paper is based on the pre-detection stellar variability filter described in \cite{AI04}. This filter is a combination of median filtering (to reduce the level of random noise), and of boxcar averaging/smoothing (to extract the long term variations: the stellar variability).

	\subsection{Deformation of the transit signal}

We tested the effect of \cite{AI04} pre-detection filter on the transit signal of 20 simulated light-curves, produced for the CoRoT blind test exercise 2 (BT2, \cite{MPBal05}). The three main components in these light-curves are the stellar variability, the transit signal, and the instrumental noise.  Figure \ref{fig1}a (bottom panel) shows a BT2 light curve of an active star with Jupiter-like transits. \\
We noticed that \cite{AI04} filter deforms the transit shape (Figure \ref{fig1}b, bottom panel). This effect is stronger for shallow transits on active stars. On average, for filtered light curves, the transit depth is under-estimated by 20\% and its duration by 15\% (compared to the value measured from the original transit with instrumental noise only). Miss-estimating the transit depth and duration leads to a miss-estimation of the planet radius, and orbital inclination (used to derive the planet mass).

\section{Iterative filtering}\label{itfilt}

We designed an iterative process to filter stellar variability and extract the original transit shape, based on \cite{AI04} filter, and using the additional post-detection knowledge on the planet orbital period.

\begin{figure}[htbp]
	\center
	 \begin{tabular}{c}
	 \includegraphics[width=9cm]{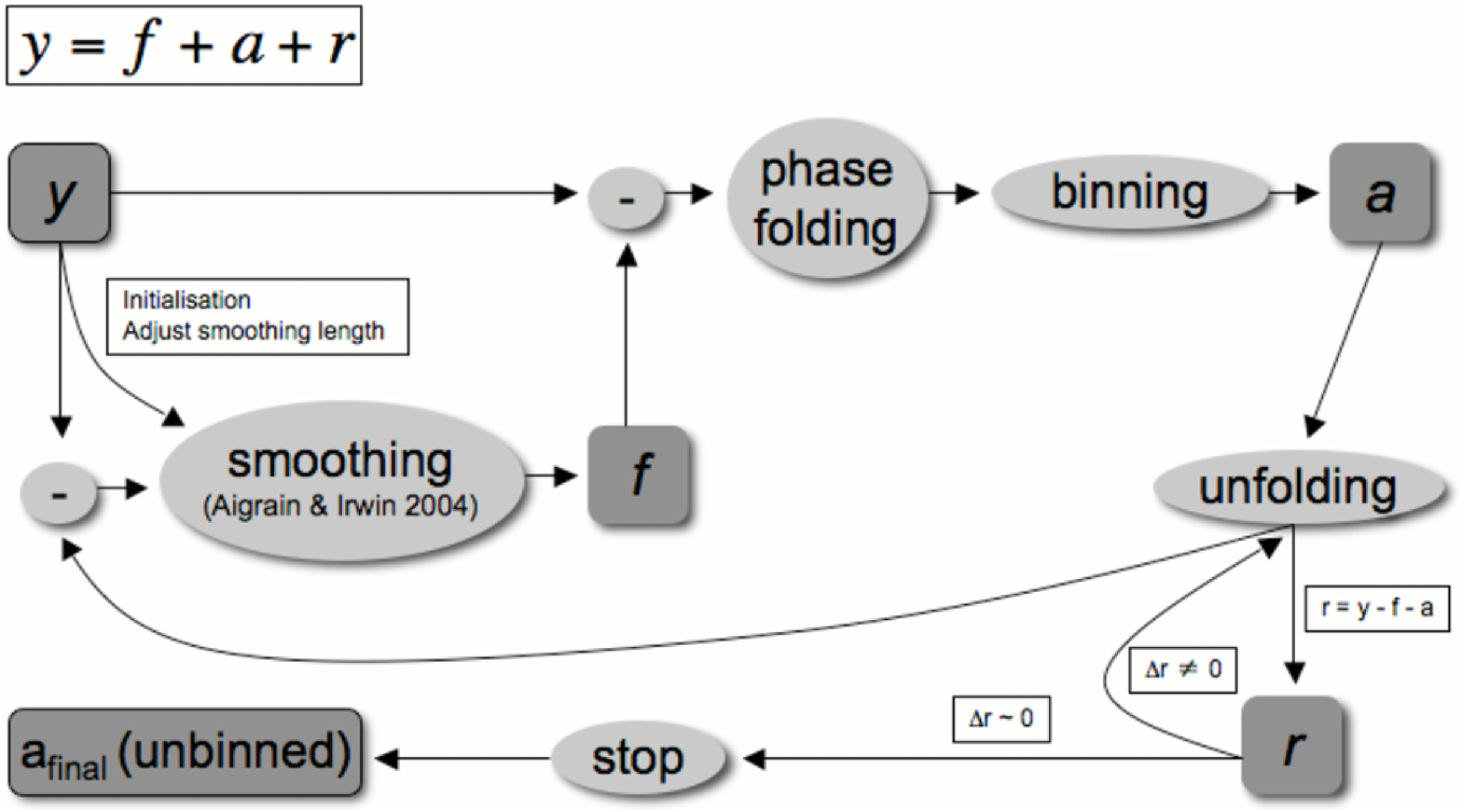}
	 \\ \\ \\
	 \includegraphics[width=6cm]{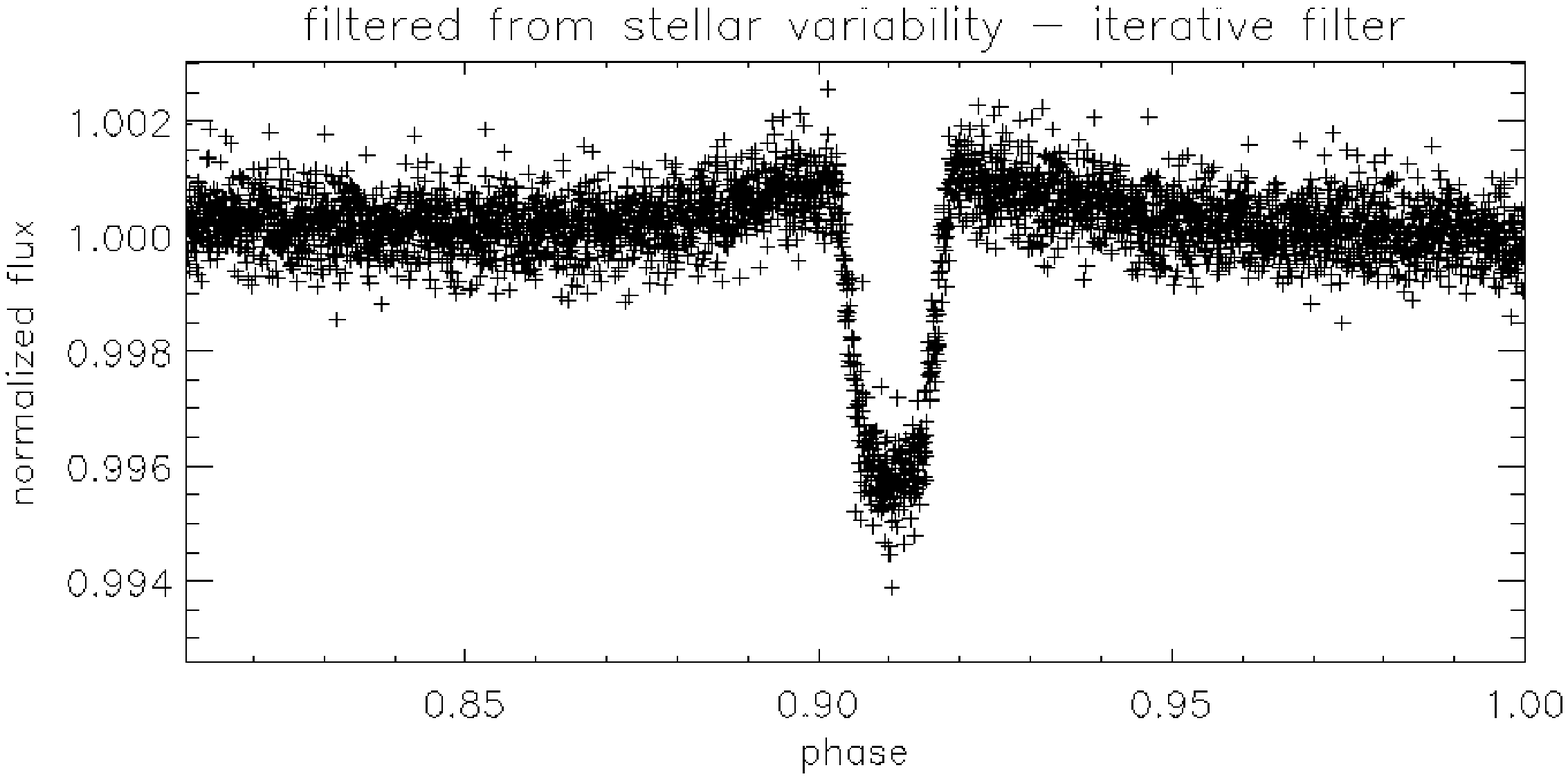}
  	\end{tabular}
	\caption{{\it Top:} Chart of the iterative filter process. The light curve ({\it y}) is decomposed into three components: the transit signal ({\it a}), the stellar variability ({\it f}), and the residual noise ({\it r}). The final estimation of the transit signal ({\it $a_{\rm final}$}) is obtained after iterating on the pre-detection filter (step called "smoothing"). {\it Bottom:} Transit ({\it $a_{\rm final}$}) resulting from the iterative filter applied to the light curve of Figure \ref{fig1}a {\it bottom plot}.}
	\label{fig2}
\end{figure}

	\subsection{Implementation of the post-detection filter \label{secImp}}

Our post-detection filter is based on \cite{KBN05} idea of decomposing the light curve into three components (transit signal, correlated noise, and white noise), to filter out systematics due to atmospheric fluctuations. In our case, we filter out stellar intrinsic variability extracted using an iterative process (top chart of Figure \ref{fig2}). \\
We estimate the stellar variability by applying \cite{AI04} pre-detection filter. We subtract the estimation from the original light curve, phase-fold the result, and bin it to reduce the random noise. We then remove the resulting transit signal from the original light curve, and start a new iteration re-evaluating the stellar variability in the light curve freed from the transit signal. We stop iterating when the residuals (original light curve minus last estimations of the transits and stellar variability) stop evolving. The best estimation of the transit signal is the original light curve minus the last estimation of the stellar variability (after 3-4 iterations).

	\subsection{Evaluation of performance}

We have applied the iterative stellar variability filter described in section \ref{secImp}, on 20 CoRoT BT2 light-curves with transits. Iterating on the estimation of the stellar variability, and using the additional knowledge on the orbital period, appears to successfully better recover the original transit shape (compare bottom of Figure \ref{fig2} with {\it bottom plot} of Figure \ref{fig1}b). On average, light curves filtered with the iterative filter gives a better estimation of the transit depth by 15\% and of the transit duration by 10\%, than when filtered with \cite{AI04} pre-detection filter.\\
The iterative filter has also been found efficient in reconstructing shallow transits that had become un-detectable after pre-detection stellar variability filtering. Another use for the iterative filter could thus be to confirm borderline transit detections in active stars.

\section{Summary and future work}

Based on 20 simulated CoRoT light curves from the BT2 light curve sample, we have shown that \cite{AI04} pre-detection filter, used to remove stellar variability prior to transit detection,  deforms the transit depth by 20\% and the transit duration by 15\%. To circumvent this, we have adapted \cite{KBN05} iterative filtering method to the case of filtering stellar variability present in space-based light curves. The resulting post-detection iterative filter improves the estimation of the transit depth and duration by 15\% and 10\% respectively. \\
The two areas where we plan to focus future efforts are {\it a)} to further automate the filtering process (some user interaction is currently needed to initialize the filter smoothing length), and {\it b)} to evaluate the improvement in the planet parameters uncertainties resulting from the improvement in the transit parameters from this work. As {\it b)} depends on the particulars of each system, we plan to derive these uncertainties using Monte Carlo simulations for known planets. Further tests will include using other pre-detection filters, including ones which can be applied to data with significant temporal gaps.

%\bibliographystyle{aa}
%\bibliography{ref.bib}

%\begin{discussion}
%\end{discussion}
    
\end{document}